\documentstyle[12pt]{article}
 \def\laq{\raise 0.4 ex \hbox{$<$}\kern -0.8 em\lower 0.62 ex\hbox{$\sim$}}
 \def\gaq{\raise 0.4 ex \hbox{$>$}\kern -0.7 em\lower 0.62 ex\hbox{$\sim$}}

\input{epsf}

\def\AJL{{\it Ap. J. Lett.} }

\def\CQG{{\it Class. Quantum Gravity} }

\def\GRG{{\it Gen. Relativity and Gravitation} }

\def\JHEP{{\it JHEP} }

\def\MNRAS{{\it Mon. Not. R. Ast. Soc.} }

\def\NC{{\it Il Nuovo Cimento} }
\def\NP{{\it Nucl. Phys.} }
\def\PL{{\it Phys. Lett.} }
\def\PR{{\it Phys. Rev.} }
\def\PRL{{\it Phys. Rev. Lett.} }

\def\al{\alpha}
\def\be{\beta}
\def\ga{\gamma}
\def\de{\delta}

\def\th{\theta}

\def\la{\lambda}

\def\De{\Delta}

\def\Om{\Omega}

 \def\frac#1#2{{\textstyle{{#1}\over
{#2}}}} 
\def\lsim{\mathrel{\rlap{\lower4pt\hbox{\hskip1pt$\sim$}}
    \raise1pt\hbox{$<$}}} \def\gsim{\mathrel{\rlap{\lower4pt\hbox{\hskip1pt$\sim$}}
    \raise1pt\hbox{$>$}}}
\def\sqr#1#2{{\vcenter{\vbox{\hrule height.#2pt
         \hbox{\vrule width.#2pt height#1pt \kern#1pt
         \vrule width.#2pt}
         \hrule height.#2pt}}}} 
\def\beq{\begin{equation}}
\def\eeq{\end{equation}}
\def\beqa{\begin{eqnarray}}
\def\eeqa{\end{eqnarray}}

\begin{document}

\begin{flushright}
{DF/IST - 5.2004} \\
{August 2004} 
\end{flushright}
\vglue 1cm

\begin{center}

{{\bf  Astrophysical Constraints on Scalar Field Models}}

\vglue 0.5cm {O.\ Bertolami and J. P\'aramos}

\vglue 0.2cm

{E-mail addresses: {\tt orfeu@cosmos.ist.utl.pt;
x\_jorge@netcabo.pt}}

\bigskip
{\it Instituto Superior T\'ecnico, Departamento de F\'\i sica,\\}
\smallskip
{\it Av. Rovisco Pais, 1049-001 Lisboa, Portugal\\}
\end{center}

\setlength{\baselineskip}{0.7cm}

\vglue 1cm

\date{\today}

\begin{abstract}
We use stellar structure dynamics arguments to extract bounds on
the relevant parameters of scalar field models: the putative
scalar field mediator of a fifth force with a Yukawa potential,
the new variable mass particle (VAMP) models, and a
phenomenologically viable model to explain the Pioneer anomaly. We
consider the polytropic gas model to estimate the effect of these
models on the hydrostatic equilibrium equation and fundamental
quantities such as the central temperature. The current bound on
the solar luminosity is used to constrain the relevant parameters
of each model.

\vskip 0.5cm

\end{abstract}

%%%%%%%%%%%%%%%%%%%%%%%%%%%%%%%%%%%%%%%%%%%%%%%%%%%%%%%%%%%%%%%%%%%%%

\section{Introduction}

Scalar fields play a crucial role in particle physics and
cosmology. Indeed, in inflation, the potential of a scalar field,
the inflaton, acts as a dynamical vacuum energy that allows for an
elegant solution of the initial conditions problems \cite{Olive}.
This prominent role of scalar fields is also evident in models to
explain the late time accelerated expansion of the Universe in
vacuum energy evolving and quintessence models
\cite{quintessence}, as well as in the Chaplygin gas dark
energy-dark matter unification model \cite{unified}. Scalar fields
have also been proposed as dark matter candidates \cite{scalar}.
Furthermore, it has been recently proposed that a scalar field can
be also at the source of the anomalous acceleration detected by
the Pioneer spacecraft \cite{Bertolami}.

Scalar fields may also have astrophysical implications as, for
instance, the mediating boson of an hypothetical fifth force,
which should yield the measurable effects on celestial bodies,
besides the other known forces of nature. Although the origin of
these fields is rather speculative, they can all be described by
the Yukawa potential, written here as $V_Y(r) = A e^{-mr} / r $,
where $A$ is the coupling strength and $m$ is the mass of the
field, which sets the range of the interaction, $\la_Y \equiv
m^{-1}$.

Models leading to a Yukawa-type potential can be found in widely
distinct areas such as braneworld models, scalar-tensor theories
of gravity and in the study of topological defects. In braneworld
models, one considers our Universe as a 3-dimensional world-sheet
embedded in a higher dimensional bulk space \cite{braneworlds}.
Symmetry considerations about the brane and its topological
properties can be implemented to constrain the evolution of matter
on the brane and gravity on the brane and in the bulk.

Braneworld models are rather trendy in cosmology and allow, for
instance, for a solution for the hierarchy problem, whether the
typical mass scale of the bulk is comparable with the electroweak
breaking scale, $M_{EW} \sim TeV$. As a result, a tower of
Kaluza-Klein (KK) massive tensorial perturbations to the metric
appears. Following the KK dimensional reduction scheme, the masses
(eigenvalues) of these gravitons (eigenfunctions) are ordered.
Most braneworld models consider one first light mode with
cosmological range, and hence all ensuing modes have sub-millimeter
range.

Given the relevance of scalar fields, the search for bounds on the
Yukawa parameters is crucial, so to exclude unviable models and
achieve some progress in the study on those that appear feasible.
Most experimental tests of a ``fifth'' force have been conducted
in the vacuum; in the authors opinion, a study on the way this
force should affect stellar equilibrium is lacking, and
constitutes one of the motivations for this work.

The bounds on parameters $A$ and $\la_Y \equiv m^{-1}$ include the
following (see \cite{Fischbach, Nunes} and references therein):
laboratory experiments devised to measure deviations from the
inverse-square law, sensitive to the range $10^{-2}~ m < \la_Y < 1
~m$, and constraining $A$ to be smaller than $10^{-4}$;
nucleosynthesis bounds which imply that $A < 4 \times 10^{-1}$ for
$ \la_Y < 1~ m$; gravimetric experiments, sensitive in the range
of $10~ m < \la_Y < 10^3~ m$, suggesting $A < 10^{-3}$; satellite
tests probing ranges of about $ 10^5~ m < \la_Y <  10^7 ~m$,
showing that $A < 10^{-5}$; and radiometric data of the Pioneer
10/11, Galileo and Ulysses spacecrafts suggesting the existence of
a new force with parameters $A = -10^{-3}$ and $\la_Y = 4 \times
10^{13}~ m$ \cite{Nieto}. It is striking that, for $\la_Y <
10^{-3}~ m$ and $\la_Y > 10^{14} ~ m$, $A$ is essentially
unconstrained. Considerations on higher dimensional superstring
motivated cosmological solutions hint that modifications to
Newtonian gravity will occur in the short range region, $\la_Y <
10^{-3} ~ m$ (cf. Figure \ref{graph_exc_y} below). This range also
emerges if one assumes the observed vacuum energy density to be
related with scalar or vector/tensor excitations \cite{Beane}.

In high energy physics, it is widely accepted that the mass of
fermions results from the Higgs mechanism, in which a scalar field
coupled to the right and left component of a particle acquires a
vacuum expectation value that acts as a mass term in the
Lagrangian density. This behaviour of the Higgs-scalar field
depends on the presence of a potential which acquires
non-vanishing minima and, therefore, cannot evolve monotonically.

Can one relax this last feature of the Higgs mechanism? This has
been the main motivation behind the variable mass particle
proposal \cite{Carroll}. In these models it is assumed that there
are some yet unknown fermions whose mass results not from a
coupling to the Higgs boson, but to a quintessence-type scalar
field with a monotonically decreasing potential. This potential
has no minima, yet the coupling of the scalar field to matter can
be included in an effective potential of the following form
$V_{eff} (\phi) = V(\phi) + \la n_\psi \phi$, where $n_\phi$ is
the number density of fermionic VAMPs and $\la$ is their Yukawa
coupling. In this way, a minimum is developed and the ensuing
vacuum expectation value (vev) is responsible for the particle's
mass. Since in a cosmological setting the density depends on the
scalar factor $a(t)$, this mass will vary on a cosmological time
scale.

Before proceeding, note that the present analogy is not perfect.
Indeed, while the Higgs mechanism relies on a spontaneous symmetry
breaking, where the vev experiences a transition from a vanishing
to a finite value, the VAMP idea assumes that \textit{no} vev
exists if the matter term of the effective potential is ``switched
off'', whereas it is always non-vanishing when the latter is
considered.

For definitiveness, we choose a potential of the quintessence-type
form, $V(\phi) = u_0 \phi^{-p}$, where $u_0$ has dimensionality
$M^{p+4}$. The effective potential $V_{eff} (\phi)$, acquires a
vev given by

\beq \phi_c \equiv < \phi > = \left({ p u_0 \over \la n_\psi}
\right)^{1/(1+p)}~~.\eeq

\noindent Since the number density evolves as $n_\psi(t) =
n_{\psi0} a(t)^{-3}$, while $\phi$ evolves as $\phi(t)=\phi_0
a^{3/(1+p)}$, where $\phi_0$ is the present value of the scalar
field, then its mass is given by

\beq m_\phi^2 \equiv \left[ {\partial ^2 V \over \partial \phi^2}
\right] _{\phi=\phi_0} = \left[p(p+1)u_0\phi_0^{-(p+2)}
\right]a^{-3(2+p)/(1+p)}~~,\eeq

\noindent and the mass of the VAMP fermions is \cite{Carroll}

\beq m_\psi = \la \phi_0 a^{3/(1+p)}~~.\eeq

Considering the evolution of the energy density contributions as a
function of the redshift, $z$, one can compute the age of the
Universe

\beq t = \int_0^a {da' \over a'} = H_0^{-1} \int_0^{(1+z^{-1})}
\left[ 1 - \Om_0 + \Om_{M0} x ^{-1} + \Om_{V} x^{(2+p)/(1+p)}
\right]^{1/2} dx~~, \eeq

\noindent where $H_0 = 100 ~h~Km~s^{-1}~Mpc^{-1}$, $\Om_{M0} $ is
the energy density of normal baryonic plus dark matter, $\Om_{V}$
is the energy density due to the potential driving $\phi$ and
$\Om_0$ is the total energy density of the Universe. The limiting
case where $\Om_0 = \Om_{V} = 1$, $\Om_{M0} = 0$ yields

\beq t_0 = {2 \over 2} H_0^{-1} \left( 1 + p^{-1} \right)~~. \eeq

Assuming that the VAMP particles, $\psi$, were relativistic when
they decoupled from thermal equilibrium, then it follows that
\cite{Carroll}

\beq m_\psi = 12.7 ~\Om_{\psi0} h^2 r_\psi^{-1} a^{3/2}~eV~~, \eeq

\noindent where $r_\psi $ is the ratio of $g_{eff}$, the effective
number of degrees of freedom of $\psi$, to $g_{*f}$, the total
effective number of relativistic degrees of freedom at freeze-out.
Furthermore, in terms of the Yukawa coupling,

\beq u_0 = 1.02 \times 10^{-9} {\Om_{\psi0}^2 h^4 \over \la
r_\psi}~eV^5~~,\eeq

\noindent and thus

\beq m_\phi = 1.00 \times 10^{-6} {\la r_\psi \over
\Om_{\psi0}^{1/2} h } a^{-9/4}~eV~~.\eeq

However interesting, VAMP models have not been subjected to a more
concrete analysis mostly due to a significant caveat, namely, the
introduction of exotic $\psi$ fermions, the VAMP particles, and
the consequent derivation of the cosmologically relevant
quantities in terms of their unknown relative density,
$\Om_{\psi0}$. As a result of this somewhat arbitrary parameter,
plus the unknown coupling constant with the scalar field and the
potential strength, there is little one can do in order to draw
definitive conclusions from VAMP models.

The present study attempts to overcome this drawback, by asserting
that, asides from the hypothetical existence of the assumed exotic
particles, all fermions couple to the quintessence scalar field.
In this assumption one considers that fermionic matter couples
mainly to the Higgs boson, so that the VAMP mass term is a small
correction to the mass acquired by the Higgs mechanism. Thus,
exotic VAMP particles are defined by their lack of the Higgs
coupling.

The extension of the VAMP proposal to usual fermionic matter
implies in a correction to the cosmologically relevant results
obtained in Refs. \cite{Carroll, Rosenfeld}. However, these
corrections should be negligible, as one assumes that the ``Higgs
to quintessence'' coupling ratio is small, that is, at
cosmological scales the exotic VAMP particles dominate the VAMP
sector of usual fermions.

Notice that the vev resulting from the effective potential depends
crucially on the particle number density $n_\psi$. Hence, it is
logical to expect that the effect of this variable mass term in a
stellar environment should be more significant than in the vacuum.
This sidesteps the model from the usual cosmological scenario with
a temporal variation, to the astrophysical case with an isotropic
spatial dependence, thus allowing one to attain bounds on model
parameters from known stellar physics observables.

Another object of this study concerns the Pioneer anomaly. This
consists in an anomalous acceleration inbound to the Sun and with
a constant magnitude of $a_A \simeq (8.5 \pm 1.3) \times 10^{-10}~
m s^{-2}$, revealed by the analysis of radiometric data from the
Pioneer 10/11, Galileo and Ulysses spacecraft. Extensive attempts
to explain this phenomena as a result of poor accounting of
thermal and mechanical effects and/or errors in the tracking
algorithms were presented, but are now commonly accepted as
unsuccessful \cite{Nieto}.

The two Pioneer spacecraft follow approximate opposite hyperbolic
trajectories away from the Solar System, while Galileo and Ulysses
describe closed orbits. Given this and recalling that one has
three geometrically distinct designs, an ``engineering'' solution
for the anomaly seems not very plausible.

Hence, although not entirely proven and even poorly understood
(see e.g. Ref. \cite{Bertolami} and references within), the
Pioneer anomaly, if a real physical phenomenon, should be the
manifestation of a new force. This force, in principle, acts upon
the Sun itself, and thus lends itself to scrutiny under the scope
of this study. It turns out that one can model its effect and
constrain, even though poorly, the only parameter involved, the
anomalous acceleration $a_A$.

\section{The polytropic gas stellar model}

Realistic stellar models, arising from the assumptions of
hydrostatic equilibrium and Newtonian gravity, rely on four
differential equations, together with appropriate definitions
\cite{Bhatia, Padmanabham}. This intricate system requires
heavy-duty numerical integration with complex code designs, and an
analysis of the perturbations induced by a variable mass is beyond
the scope of the present study. Instead, we focus the polytropic
gas model for stellar structure: this assumes an equation of state
of the form $P = K \rho^{n+1/n}$, where $n$ is the so-called
polytropic index, that defines intermediate cases between
isothermic and adiabatic thermodynamical processes, and $K$ is the
polytropic constant, defined below. This assumption leads to
several scaling laws for the relevant thermodynamical quantities,

\beqa \rho = \rho_c \th^n(\xi)~~,~~~~(a) \\ \nonumber T = T_c
\th(\xi)~~,~~~~(b) \\ P = P_c \th(\xi)^{n+1}~~,~~~~(c)
\nonumber\eeqa

\noindent where $\rho_c$, $T_c$ and $P_c$ are the values of the
density, temperature and pressure at the center of the star.

The function $\th$, responsible for the scaling of $P$, $\rho$ and
$T$, is a dimensionless function of the dimensionless variable
$\xi$, related to the physical distance to the star's center by $r
= \al \xi$, where

\beq \al = \left[{(n+1)K \over 4 \pi
G}\rho_c^{(1-n)/n}\right]^{1/2} ~~, \eeq

\beq K = N_n G M^{(n-1)/n} R^{(3-n)/n} ~~ \label{k} \eeq

\noindent and

\beq N_n = \left[{n+1 \over (4\pi)^{1/n}} \xi^{(3-n)/n}
\left(-\xi^2 {d \th \over d \xi}
\right)^{(n-1)/n)}\right]^{-1}_{\xi_1}~~,\eeq

\noindent so that $R$ is the star's radius, $M$ its mass and
$\xi_1$, defined by $\th(\xi_1) \equiv 0$, corresponds to the
surface of the star (actually, this definition states that all
quantities tend to zero as one approaches the surface. Its
unperturbed value is $\xi_{1}^{(0)} = 6.89685$, as given in Ref.
\cite{Bhatia}.

The function $\th(\xi)$ obeys a differential equation arising from
the hydrostatic equilibrium condition

\beq {d \over dr} \left( {dP \over dr}{r^2 \over \rho} \right) = -
G {dM(r) \over dr} ~~, \eeq

\noindent the Lane-Emden equation:

\beq {1 \over \xi^2} {\partial \over \partial \xi} \left(\xi^2
{\partial \th \over \partial \xi} \right) = -\th^n~~. \eeq

\noindent We point out that the physical radius and mass of a star
appears only in Eq. (\ref{k}) so that the behaviour of the scaling
function $\th(\xi)$ is unaffected by their values. Hence, the
stability of a star is independent of its size or mass, and
different types of stars correspond to different polytropic
indices $n$. This kind of scale-independent behavior is related to
the homology symmetry of the Lane-Emden equation.

The first solar model ever considered corresponds to a polytropic
star with $n=3$ and was studied by Eddington in 1926. Although
somewhat incomplete, this simplified model gives rise to relevant
constraints on the physical quantities.

\vskip 3cm

\section{Results}

\vskip 0.5cm {\bf 3.1 Yukawa potential induced perturbation}

In this subsection we look at the hydrostatic equilibrium equation
with a Yukawa potential:

\beq dP = - {G M(r) \left[1 + A e^{-mr}\right] \over r^2} \rho(r)
dr ~~. \eeq

\noindent which, after a small algebraic manipulation, implies
that

\beq  {1 \over r^2} {d \over dr} \left({dP \over dr} {r^2 \over
\rho} \right) = -4 \pi G \rho \left[1 + Ae^{-mr}\right] + {G M(r)
Am e^{-mr} \over r^2} ~~. \label{hydro_yp} \eeq

\noindent The last term is a perturbation to the usual Lane-Emden
equation, obtained by substituting $ r = \al \xi $ and $ \rho =
\rho_c \th^n$; one gets

\beq {1 \over \xi^2} {d \over d \xi} \left( \xi^2 {d \th \over d
\xi}\right) = -\th^n \left[1 + Ae^{-m \al \xi}\right] + {M(\xi) A
m e^{-\al m \xi} \over 4 \pi \rho_c \al^2 \xi^2} ~~.\eeq

\noindent Since $ \al = \left({(n+1)K \over 4 \pi
G}\rho_x^{(1-n)/n}\right)^{1/2} $ and $M(\xi) = -4 \pi
\left[{(n+1)K \over 4 \pi G}\right]^{3/2} \rho_c^{(3-n)/2n} \xi^2
{d\th \over d\xi}$, the second perturbation term can be written as

\beq {M(\xi) Am e^{-\al m \xi} \over 4 \pi \rho_c \al^2 \xi^2} = -
\left[{(n+1) K \over 4 \pi G} \right]^{1/2} \rho_c^{(1-n)/2n} {d
\th \over d\xi} Am e^{-\al m \xi} ~~. \label{pert_yp} \eeq

\noindent Furthermore, one has

\beqa K & = & N_n G M^{(n-1)/n} R^{(3-n)/n}~~, \\ \nonumber P_c &
= & W_n G M^2 / R^4~~, \\ \nonumber P_c & = & K
\rho_c^{(n+1)/n}~~, \eeqa

\noindent where $R$, $M$ is the radius and mass of the star and
$N_n$, $W_n$ are numbers which depend on $n$, and for which one
takes the tabulated values, valid for the unperturbed equation.
Hence,

\beq \rho_c = \left({W_n \over N_n}\right)^{(1-n)/2(1+n)} \left({M
\over R^3}\right)^{(1-n)/2n}~~.\eeq

\noindent Substituting into Eq.(\ref{pert_yp}), one obtains

\beq -\left[{(n+1) K \over 4 \pi G} \right]^{1/2}
\rho_c^{(1-n)/2n} {d \th \over  d\xi} Am e^{-\al m \xi} =
-\sqrt{{n+1 \over 4 \pi}} N_n^{n/(n+1)} W_n^{(1-n)/2(n+1)} R {d
\th \over d \xi} Am e^{-\al m \xi}~~. \eeq

\noindent If one now defines the dimensionless quantities

\beq C_n \equiv \left({n+1 \over 4 \pi}\right)^{1/2} N_n^{n/(n+1)}
W_n^{(1-n)/2(n+1)}~~, ~~ \ga \equiv m R~~, \eeq

\noindent the perturbed Lane-Emden equation acquires the form

\beq {1 \over \xi^2} {d \over d \xi} {d \th \over d \xi}  = -\th^n
\left[1 + Ae^{-\al m \xi}\right] - \ga A C_n {d \th \over d \xi}
e^{-\al m \xi} ~~.\eeq

\noindent One can eliminate $\al$ on the exponential term by
writing $ -\al m \xi = -\al m R \xi / R = -\ga \xi / \xi_s $,
where $\xi_s$ corresponds to the value at the surface of the star.
Since one must specify it prior to integration of the differential
equation, one assumes that $\xi_s \simeq \xi_1$, the latter being
the tabulated value for $\th(\xi_1)=0$. This is in good
approximation, since the solar temperature is very small when
compared at the surface to its central value, $T_s = 5.778 \times
10^3 ~K = 3.7 \times 10^{-4} T_c$. Hence, one gets

\beq {1 \over \xi^2} {d \over d \xi} {d \th \over d \xi}  = -\th^n
\left[1 + Ae^{-\ga \xi / \xi_1}\right] - \ga A C_n {d \th \over d
\xi} e^{- \ga \xi / \xi_1} ~~.\eeq

This notation makes clear that the perturbation vanishes for $A
\rightarrow 0$. Notice that, if $m \sim R^{-1}$, then $\ga \sim
1$. However, since $m$ should arise from a fundamental theory,
this would be the case only for stars of a particular size $m^{-1}
\equiv \la_Y $. Since $\ga \equiv m R = R / \la_Y $, it is clear
that large stars ($R \gg \la_Y \rightarrow \ga \gg 1 $) are
perturbed only within a small central region, where $\xi / \xi_1
\ll 1$.

It is also apparent that the Yukawa-type perturbation breaks the
invariance under homologous transformations, since $\ga = mR$
explicitly depends on $R$, the radius of the star. Hence, a bound
on $\ga$ obtained from the luminosity or other observables is, for
a given $m$, equivalent to a bound on the maximum size of a
polytropic star of index $n$.

The boundary conditions are unaffected by the perturbation: from
the definition $\rho = \rho_c \th(\xi)$, one gets $\th(0)=1$; the
hydrostatic equation Eq. (\ref{hydro_yp}), in the limit $\xi
\rightarrow 0$, still imposes that $\left|{d \th / d \xi}
\right|_{\xi=0} = 0$.

\begin{figure}

\epsfysize=10cm \epsffile{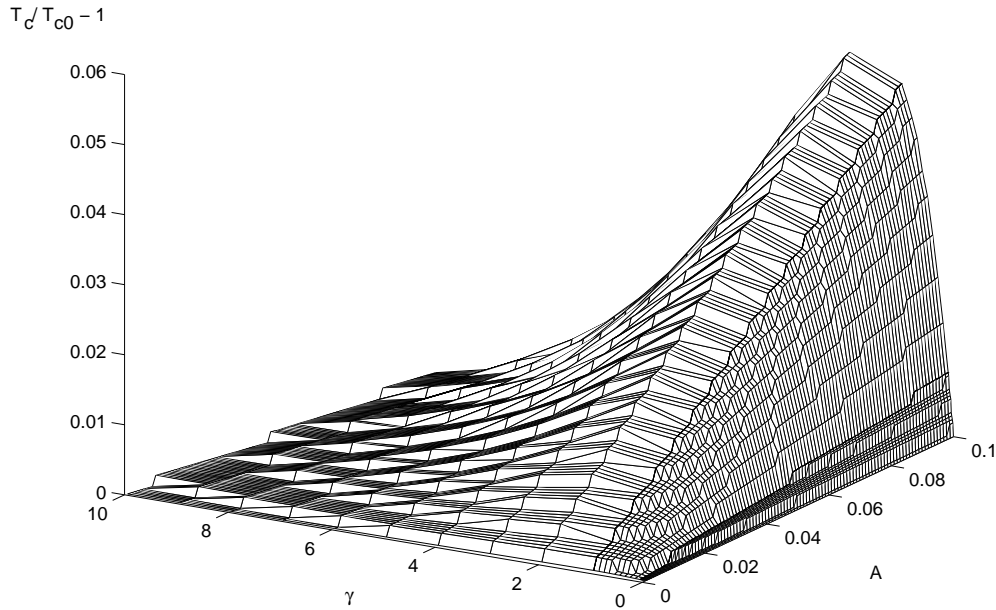} \caption{Relative
deviation from unperturbed central temperature $T_c/T_{c0}-1$, for
$A$ ranging from $10^{-3}$ to $10^{-1}$, and $\ga$ from $10^{-1}$
to $10$.} \label{graph_Tc_y}

\end{figure}

By specifying $A$ and $\ga$, we may solve Eq. (\ref{hydro_yp})
numerically and get relative deviations from the unperturbed
solution. Since a star's central temperature is given as a
function of $M$ and $R$ and in terms of the mean molecular weight
$\mu$, the Hydrogen mass $H$ and the Boltzmann constant $k$ by $
T_c = Y_n \mu GM/R$, with

\beq Y_n = \left({3 \over 4}\right)^{1/n} (H/k) N_n \left[-{\xi
\over 3 {d \th \over d \xi}} \right]^{1/n}_{\xi_1}~~, \eeq

\noindent one can compute the relative changes on $T_c$ for
different values of $A$ and $m$. The results are presented in
Figure \ref{graph_Tc_y}. The parameters were chosen so that the
Yukawa interaction $\la_Y$ ranges from $0.1R$ to $10R$: in the
first case, the interaction is mainly located in the interior of
the star, while in the second case it reaches outward and could be
considered approximately constant within it. The Yukawa coupling
was chosen so that the effect on $T_c$ could be sizeable (that is,
of order $O(10^{-4})$).

%%%%%%%%%%%%%%%%%%%%%%%%%%%%%%%%%%%%%%%%%%%%%%%%%%%%
%%%%%%%%%%%%%%%%%%%%%%%%%%%%%%%%%%%%%%%%%%%%%%%%%%%%

\begin{figure}

\epsfysize=10cm \epsffile{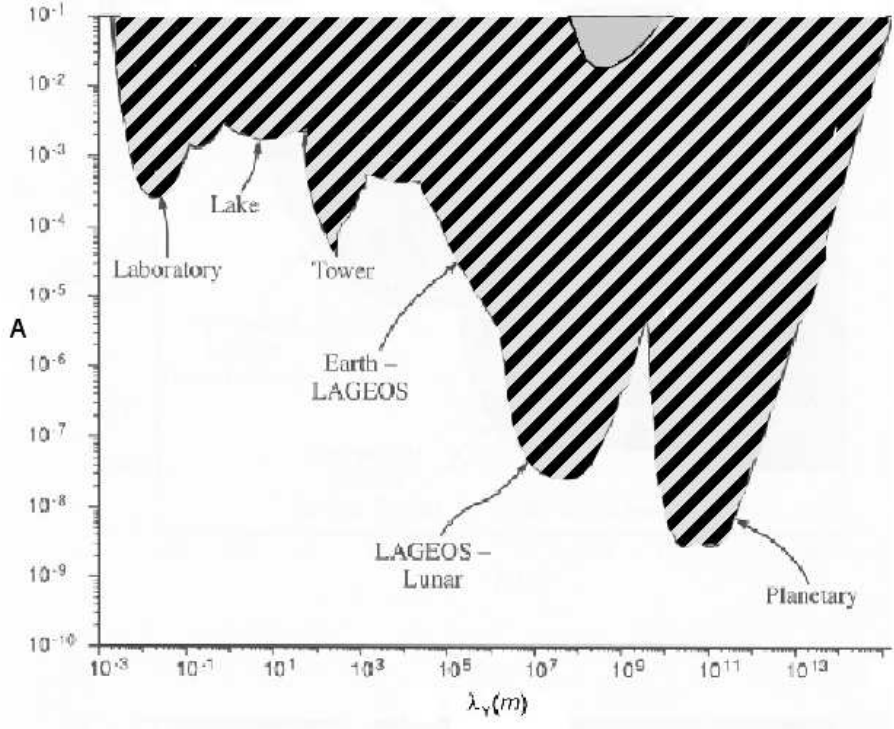} \caption{Exclusion plot
for the relative deviation from unperturbed central temperature
$T_c$, for $A$ ranging from $10^{-3}$ to $10^{-1}$, and $\ga$ from
$10^{-1}$ to $10$ (tip at the top), superimposed on the available
bounds \cite{Fischbach}} \label{graph_exc_y}

\end{figure}

This enables us to build the exclusion plot of Figure
\ref{graph_exc_y}, by imposing that $\De T_c < 4 \times 10^{-3}$,
the accepted bound derived from Solar luminosity constraints
\cite{Bhatia}. It is superimposed on the different bounds
available \cite{Fischbach}.

%%%%%%%%%%%%%%%%%%%%%%%%%%%%%%%%%%%%%%%%%%%%%%%%%%%%
%%%%%%%%%%%%%%%%%%%%%%%%%%%%%%%%%%%%%%%%%%%%%%%%%%%%

\vskip 0.5cm {\bf 3.2 VAMP models}

As previously discussed, we consider usual fermions with a
variable mass term $\de m$ given by the coupling to the
quintessence-type scalar field; its mass is then given by the
usual Higgs-mechanism related term $m$ plus this new VAMP term, a
small contribution: $m = m_{Higgs} + \de m$. We adopt, however, a
``worst case'' scenario, in which the electron mass is not mainly
due by the Higgs mechanism plus a minor VAMP sector contribution,
but fully given by the said VAMP component alone. This implies
that the cosmological expectation value should be weakly
perturbed, so that the electron mass does not undergo large
variations.

It can be shown (see Appendix) that this variable term leads to a
geodesic deviation equation of the form

\beq \ddot{x}^a = \left[ \Gamma^a_{bc} + {\al_a \over 2 \al}
g_{bc} \right] \dot{x}^b \dot{x}^c + {2 \dot{\al} \over \al}
\dot{x}^a ~~.\eeq

\noindent Assuming isotropy, one obtains the acceleration

\beq \vec{a} = \vec{a}_{Newton} + {m' \over m} g_{bc} \dot{x}^b
\dot{x}^c \vec{u}_r - {\dot{m} \over m} \vec{v} = \vec{a}_{Newton}
+ {\phi' \over \phi} g_{bc} \dot{x}^b \dot{x}^c \vec{u}_r -
{\dot{\phi} \over \phi} \vec{v}~~. \eeq

\noindent where the prime and the dot denote derivatives with
respect to $r$ and $t$, respectively. Considering the Newtonian
limit $ g_{ab} = diag(1,-1,-1,-,1) $ so that $ g_{bc} \dot{x}^b
\dot{x}^c = 1- v^2 \simeq 1$, one finds a radial, anomalous
acceleration plus a time-dependent drag force:

\beq a_A = {\phi' \over \phi} < 0, ~~ a_D = -{\dot{\phi} \over
\phi} < 0~~. \eeq

\noindent Notice that this radial Sun bound acceleration has the
qualitative features for a possible anomalous acceleration
measured by the Pioneer probes \cite{Nieto}.

The time-dependent component should vary on cosmological
timescales, and can thus can be absorbed in the usual Higgs mass
term. Hence, one considers only the perturbation to the Lane-Emden
equation given by the radial force, $a_A \simeq \phi' / \phi$:

\beq dP = \left[ -G M(r) + a_A r^2 \right] {\rho dr \over r^2}~~,
\eeq

\noindent which translates into

\beq {1 \over r^2} {d \over dr} \left[ {r^2 \over \rho} {dP \over
dr} \right] = -4 \pi G \rho - {c^2 \over \phi(r)} \left[{2
\phi'(r) \over r} + \phi''(r) - {\phi'^2(r) \over \phi(r) }
\right] ~~. \label{hydro_vamp} \eeq

\noindent Defining the dimensionless quantities

\beqa U & \equiv & {G M \over R c^2} = 2.12 \times 10^{-6} ~~, \\
\nonumber C_n^{-1} & \equiv & (n+1) N_n^{n/(n+1)} W_n^{1 /
(n+1)}~~, \eeqa

\noindent where

\beq W_n = {1 \over 4 \pi (n+1)\left({d \th \over d \xi}
\right)^2_{\xi_1}}~~, \eeq

\noindent one obtains the perturbed Lane-Emden equation:

\beq {1 \over \xi^2} {d \over d \xi} \left[ \xi^2 {d \th \over d
\xi}\right] = - \th^n(\xi) - {C_n \over U} {1 \over \phi(\xi)}
\left[\phi''(\xi) + {2 \over \xi} \phi'(\xi) - {\phi'^2(\xi) \over
\phi(\xi) } \right]~~. \label{Lane-Emden} \eeq

The Klein-Gordon type equation for the scalar field, written in
terms of the $\xi$ variable is, inside the star, given by

\beq {1 \over \al^2} \left[\phi''(\xi) + {2 \over \xi} \phi'(\xi)
\right] = -p u_0 \phi^{-(p+1)}(\xi) + {\la \rho_c \th^n(\xi) \over
\mu}~~, \eeq

\noindent where now the prime denotes derivation with respect to
$\xi$. The parameter $\mu$ is the mean molecular weight and it is
assumed that there is one electron per molecule, that is, that the
star is composed by ``Hydrogen'' with a molecular weight $\mu$.

Beyond the star, in the Klein-Gordon equation one has the coupling
to the constant number density of fermions in the vacuum $n_\psi
\equiv n_V = 3 ~m^{-3} $,

\beq {1 \over \al^2} \left[\phi''(\xi) + {2 \over \xi} \phi'(\xi)
\right] = -p u_0 \phi^{-(p+1)}(\xi) + {\la n_V}~~. \eeq

These equations constitute a set of coupled differential equations
for $\phi(\xi)$ and $\th(\xi)$ and, of course, the continuity of
$\phi$ across the surface of the star must be addressed. A
complete derivation can be found in Ref. \cite{Bertolami}. In
here, $\al$ and $\rho_c$ depend on $W_n$, $N_n$ and related
quantities, which are evaluated after the solution $\th(\xi)$ is
known, together with $M$ and $R$. Hence, one considers their
unperturbed values for the Sun: $\al = R / \xi_1^{(0)} = 1.009
\times 10^8~m$ and $\rho_c = 1.622 \times 10^5~kg~m^{-3} $.
Moreover, current solar estimatives indicate that $\mu \simeq
0.62~m_p$, $m_p$ being the Hydrogen atomic mass \cite{Bhatia}.

For simplicity, we deal only with the case of $p=1$, as in Ref.
\cite{Carroll}. Instead of the potential strength $u_0$, we work
with the potential energy density $\Om_{V} < 1$ of the scalar
field. Before presenting the obtained numerical solutions, we
develop the expression for $\Om_V = V(\phi_c) / \rho_{crit}$, with
$\rho_{crit} \simeq 1.88 \times 10^{-29}~h^2~g~cm^{-3}$; in what
follows we chose $ h=0.71 $. Therefore

\beq V(\phi_c) = u_0 \left( \sqrt{{ u_0 \over \la n_V}}
\right)^{-1} + \la n_V \sqrt{ u_0 \over \la n_V} = 2\sqrt{u_0 \la
n_V} ~~, \eeq

\noindent which implies

\beq u_0 = {\Om_V^2 \rho_{crit}^2 \over 4 \la n_V}~~.\eeq

We now rescale the scalar field so to work with a dimensionless
quantity $\Phi \equiv \phi / \phi_c^*$, where $\phi_c^*$ is the
cosmological vev obtained by assuming as reference values $\la =
\Om_V = 1$,

\beq \phi_c^* = {\rho_{crit} \over 2 n_V}~~. \eeq

\noindent Hence, the cosmological vev for general $\la$ and $u_0$
is given by

\beq \phi_c = {\Om_V \rho_{crit} \over 2 \la n_V} = {\Om_V \over
\la} \phi_c^*~~. \eeq

\noindent so that $\Phi_c = \phi_c/\phi_c^* = \Om_V / \la$. Thus,
the Klein-Gordon equation has the following form:

\vskip 0.5cm

\noindent i) Inside the star

\beq \Phi''(\xi) + {2 \over \xi} \Phi'(\xi) = - {2 \al^2 n_V^2
\Om_V^2 \over \rho_{crit} \la} \Phi^{-2}(\xi) + {2 \al^2 \la n_V
\over \rho_{crit}} {\rho_c \over \mu } \th^3(\xi)~~,
\label{scalarint} \eeq

\noindent ii) In the vacuum

\beq \Phi''(\xi) + {2 \over \xi} \Phi'(\xi) = - {2 \al^2 n_V^2
\Om_V^2 \over \rho_{crit} \la} \Phi^{-2}(\xi) + {2 \al^2 \la n_V
\over \rho_{crit}} n_V ~~. \label{scalarext} \eeq

The perturbed Lane-Emden equation assumes the form

\beq {1 \over \xi^2} {d \over d \xi} \left[ \xi^2 {d \th \over d
\xi}\right] = - \th^n(\xi) - {C_n \over U} {1 \over \Phi(\xi)}
\left[\Phi''(\xi) + {2 \over \xi} \Phi'(\xi) - {\Phi'^2(\xi) \over
\Phi(\xi) } \right]~~. \label{Lane-Emden-Phi} \eeq

\vskip 0.5cm

Since the perturbation on $\th(\xi)$ is to be shown to be small,
one can take the unperturbed function $ \th(\xi) \simeq
\th_0(\xi)$ when solving Eqs. (\ref{scalarint}) and
(\ref{scalarext}), and then introduce the obtained solution for
the scalar field $\Phi(\xi)$ in Eq. (\ref{Lane-Emden-Phi}). First,
one assumes that the scalar field is given by its cosmological vev
perturbed by a small ``astrophysical'', $\Phi_a(\xi)$,
contribution, $\Phi(\xi) = \Om_V / \la + \Phi_a(\xi)$. Hence, the
Klein-Gordon equation becomes

\beq \Phi''_a(\xi) + {2 \over \xi} \Phi'_a(\xi) \approx {2 \al^2
\la n_V \over \rho_{crit}} {\rho_c \over \mu } \th^3(\xi) -{{2
\al^2 \la^2
 n_V^2 \over \rho_{crit}}} \left[1-{ 2 \la \Phi_a(\xi) \over \Om_V} \right] ~~,
\eeq

\noindent inside the star, and

\beq \Phi''_a(\xi) + {2 \over \xi} \Phi'_a(\xi) \approx {2 \al^2
\la n_V^2 \over \rho_{crit}} -{{2 \al^2 \la^2 n_V^2 \over
\rho_{crit}}} \left[1-{ 2 \la \Phi_a(\xi) \over \Om_V} \right] ~~,
\eeq

\noindent in the outer region.

Substituting by the Sun values $\al = R / \xi_1^{(0)} = 1.009
\times 10^8~m$, $\rho_c = 1.622 \times 10^5~kg~m^{-3} $, $\mu
\approx 0.62~m_p$, $n_\psi = 3~m^{-3}$, one gets

\beq \Phi''_a(\xi) + {2 \over \xi} \Phi'_a(\xi) \approx 6.8 \la
\left[ 4.79 \times 10^{32} \th^3(\xi) - \la \left(1-{ 2 \la
\Phi_a(\xi) \over \Om_V} \right) \right] ~~, \eeq

\noindent inside the star, and

\beq \Phi''_a(\xi) + {2 \over \xi} \Phi'_a(\xi) \approx 6.8 \la
\left[1 - \la \left(1-{ 2 \la \Phi_a(\xi) \over \Om_V}
\right)\right] ~~, \eeq

\noindent in the outer region.

Numerical integration of these equations enables the computation
of the central temperature's relative deviation; as boundary
conditions for $\Phi(\xi)$ it is imposed that both the field and
its derivative vanish beyond the Solar System (about $10^5~ AU$).
One can see by inspection that the solution $\Phi_a(\xi)$ is
practically the same, regardless of the value for $\Om_V$, as one
always assumes $\Phi_a(\xi) \ll \Om_V / \la$. However, it is
highly sensitive to $\la$.

Also, one must verify the validity of the condition $\Phi_a \ll
\Om_V / \la$ for chosen $\Om_V$ and $\la$ values. For this, note
that $\Phi_a(\xi)$ evolves as $\la^{-1}$ and, as stated above, it
is fairly independent of $\Om_V$ to a very good approximation.
Hence, choosing a smaller value for $\Om_V$ amounts to reducing
$\la$, both by lowering the field $\Phi_a(\xi)$ and increasing its
upper limit, $\Om_V / \la$.

\begin{figure}

\epsfysize=10cm \epsffile{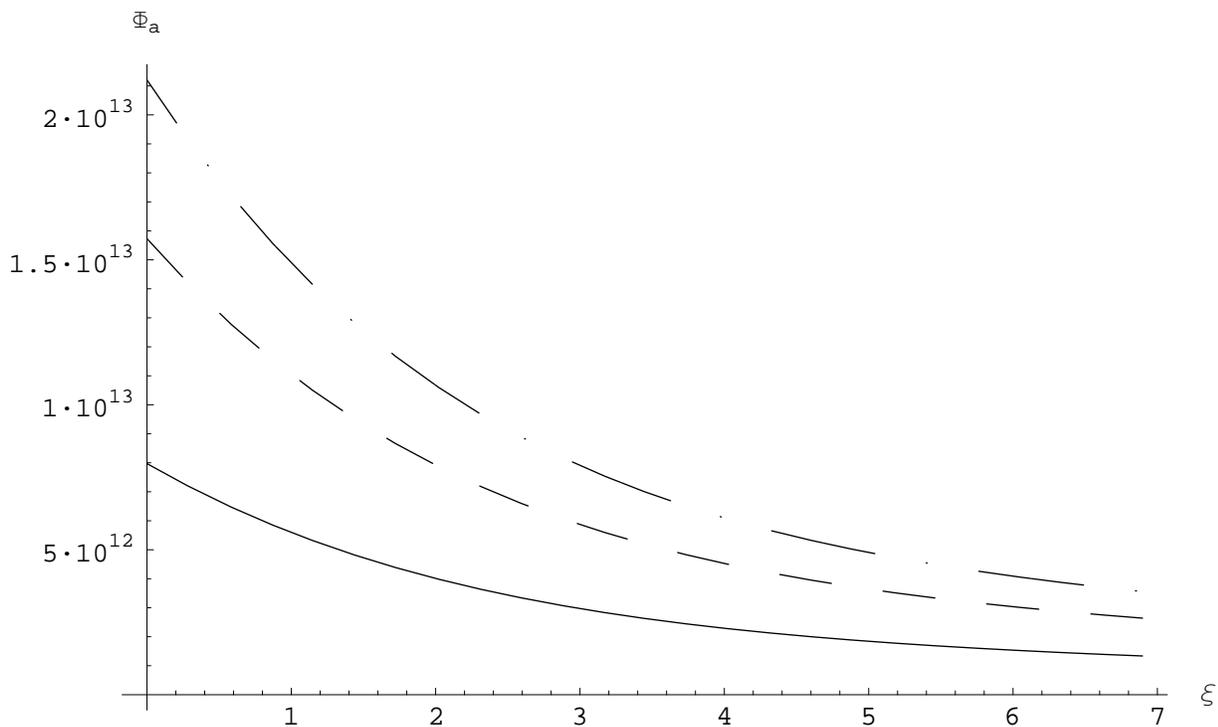} \caption{$\Phi_a(\xi)$ field
profile, for the ($\Om_V = 0.1$, $\la = 1.24 \times 10^{-14}$)
(solid line), ($\Om_V = 0.4$ , $\la = 2.45 \times 10^{-15}$)
(dashed line) and ($\Om_V = 0.7$ and $\la = 3.3 \times 10^{-14}$)
(dash-dotted line) cases.} \label{graph_phia_v}

\end{figure}

\begin{figure}

\epsfysize=7cm \epsffile{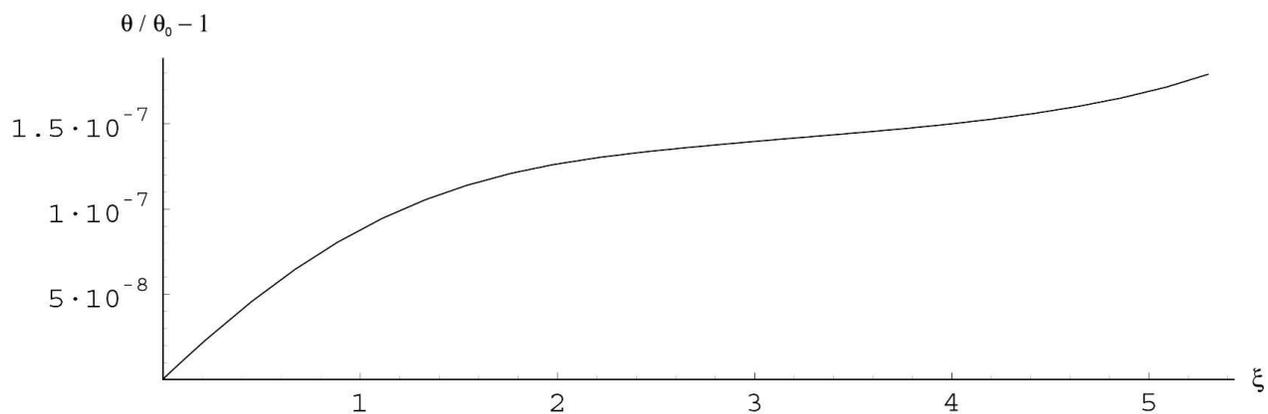} \caption{Perturbed solutions
for $\th(\xi) \sim T(\xi)$, for the $\Om_V = 0.1$, $\la = 1.06
\times 10^{-14}$ case; other solutions overlap.}
\label{graph_theta_v}

\end{figure}

By the same token, each value of $\Om_V $ corresponds a maximum
allowed value for the coupling, $\la_{max}(\Om_V)$. One then uses
these values to numerically obtain to first order solutions for
$\th(\xi)$ and $\phi(\xi)$, for say $\Om_V=0.1$, $0.4$ and $0.7$,
as presented on Figures \ref{graph_phia_v} and
\ref{graph_theta_v}. This enables one to extract the variation of
the central temperature, $T_c$. The maximum allowed values for
$\la$ (depending on the chosen $\Om_V$) are: for $\Om_V=0.1$,
$\la_{max} = 1.24 \times 10^{-14}$; for $\Om_V=0.4$, $\la\leq 2.45
\times 10^{-14}$ and for $\Om_V=0.7$, $\la\leq 3.3 \times
10^{-14}$. The limiting case $\Om_V=1$ yields $\la \leq 3.93
\times 10^{-14}$.

%%%%%%%%%%%%%%%%%%%%%%%%%%%%%%%%%%%%%%%%%%%%%%%%%%%%
%%%%%%%%%%%%%%%%%%%%%%%%%%%%%%%%%%%%%%%%%%%%%%%%%%%%

%%%%%%%%%%%%%%%%%%%%%%%%%%%%%%%%%%%%%%%%%%%%%%%%%%%%
%%%%%%%%%%%%%%%%%%%%%%%%%%%%%%%%%%%%%%%%%%%%%%%%%%%%

None of the presented curves exceed the maximum allowed variation
for $T_c$ of $0.4 \%$: the maximum of $\de T_c = 2.82 \times
10^{-8}$ occurs for $\Om_V=0.7$, $\la=2.82 \times 10^{-14}$.
Hence, the luminosity constraint is always respected and the bound
one must respect is $\la < 10^{-14}$.

\vskip 0.5cm {\bf 3.3 The Pioneer anomaly}

Following the method outlined above, we look at the hydrostatic
equilibrium equation for a constant perturbation:

\beq dP = \left[ -G M(r) + a_A r^2 \right] {\rho dr \over r^2} ~~,
\eeq

\noindent and therefore

\beq {1 \over r^2} {d \over dr} \left[ {r^2 \over \rho} {dP \over
dr} \right] = -4 \pi G \rho + {2 a_A \over r} ~~. \label{hydro}
\eeq

The last term is a perturbation to the usual Lane-Emden equation,
which is given by

\beq {1 \over \xi^2} {d \over d \xi} \left[ \xi^2 {d \th \over d
\xi}\right] = - \th^n(\xi) + {a_A \over 2 \pi G \rho_c \al \xi}
~~.\eeq

The factor in the perturbation term can be written as

\beq 2 \pi G \rho_c \al = \left[(n+1) G K \pi \right]^{1/2}
\rho_c^{(n+1)/2n} ~~. \label{pert_a} \eeq

Following the same steps as before, and defining the dimensionless
quantities

\beq C_n^{-1} \equiv \sqrt{(n+1)\pi W_n}, ~~ \be \equiv {a_A R^2
\over GM} \equiv {a_A \over a_\odot}= 3.65 \times 10^{-3}~a_A~~,
\eeq

\noindent one obtains the perturbed Lane-Emden equation

\beq {1 \over \xi^2} {d \over d \xi} \left[ \xi^2 {d \th \over d
\xi}\right] = - \th^n(\xi) + \be C_n {1 \over \xi} ~~.\eeq

\begin{figure}

\epsfysize=10cm \epsffile{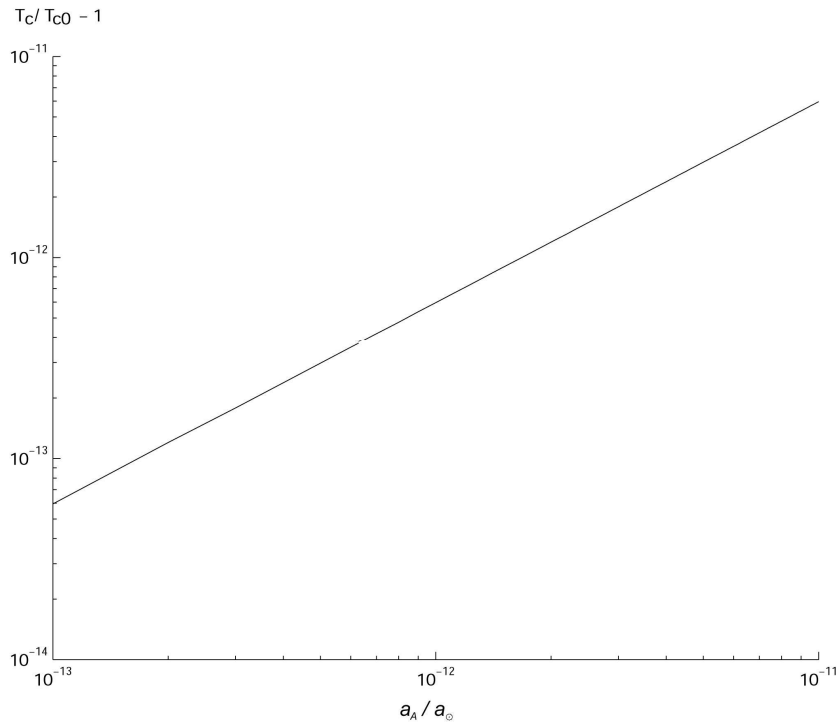} \caption{Relative deviation
from the unperturbed central temperature, for $a_A$ ranging from
$10^{-13}a_\odot$ to $10^{-11}a_\odot$.} \label{graph_Tc_a}

\end{figure}

As previously, the boundary conditions for this modified
Lane-Emden equation are unaffected by the perturbation: from the
definition $\rho = \rho_c \th(\xi)$, one gets $\th(0)=1$; the
hydrostatic equation Eq. (\ref{hydro}), in the limit $\xi
\rightarrow 0$, still imposes $\left|{d \th / d \xi}
\right|_{\xi=0} = 0$. In the present case one has only one model
parameter, $\be$, which can be constrained by the same luminosity
bounds as before. Solutions for this equation with $\be$ varying
from $10^{-13}a_\odot$ to $10^{-11}a_\odot$ (the reported value is
of magnitude $a_{P} \sim 3 \times 10^{-12} a_\odot$) enable one to
compute the relative central temperature deviation as a function
of $\be$, as presented in Figure \ref{graph_Tc_a}.

%%%%%%%%%%%%%%%%%%%%%%%%%%%%%%%%%%%%%%%%%%%%%%%%%%%
%%%%%%%%%%%%%%%%%%%%%%%%%%%%%%%%%%%%%%%%%%%%%%%%%%%%

From Figure \ref{graph_Tc_a}, one concludes that the relative
deviation of the central temperature scales linearly with $a_A$,
as $\de T_c \sim a_A/a_\odot$. Thus, the bound $\de T_c <4 \times
10^{-3}$ is satisfied for values of this constant anomalous
acceleration up to $a_{Max} \sim 10^{-4} a_\odot$. The reported
value is then well within the allowed region, and has a negligible
impact on the astrophysics of the Sun.

\section{Conclusions}

In this work we have studied solutions of the perturbed Lane-Emden
equation for three diferent cases, related to relevant scalar
field models. We obtain bounds on the parameter space of each
model from Solar luminosity constraints.

The exclusion plot obtained for a Yukawa perturbation produces no
new exclusion region in the parameter space $A$-$\la_Y$. This
results from the low accuracy to which the central temperature
$T_c$ is known, when compared to the sensibility of dedicated
experiments \cite{Fischbach}. Therefore, it is fair to expect that
improvements in the knowledge of the Sun's central temperature
could yield a new way of exploring the available range of
parameters.

For the VAMP case we have shown that the Yukawa coupling of the
VAMP sector is constrained to be $\la<10^{-14}$, and that the
Solar luminosity constraint is always respected. The numerical
analysis reveals that $\Om_V$ and $\la$ should satisfy the
relation $\la / \Om_V < 10^{-13}$.

It has also been shown that the scalar field acquires its
``cosmological'' value just outside the star, leading to no
differential shifts of the particle masses in the vacuum; thus,
there is no observable variation of fermionic masses, and hence no
violation of the Weak Equivalence Principle.

Finally, we found that an anomalous, constant acceleration such as
the one reported on the Pioneer 10/11 spacecrafts is allowed
within the Sun for values up to $10^{-4}a_\odot$, thus clearly
stating that the observed value $a_{P} \sim 10^{-12} a_\odot$ has
negligible impact on the central temperature and other stellar
parameters.

\section{Appendix}

In order to encompass models with a variable mass, we consider the
generalized Lagrangian density

\beq \textsl{L}=\sqrt{\al(x)}\sqrt{g_{ab} \dot{x}^a \dot{x}^b}~~.
\eeq

\noindent The function $\al$ is, in the homogeneous and
time-independent case, identified with the square of the rest
mass. We now deduce the Euler-Lagrange equation for the timelike
geodesics. Notice that there is no right side terms because $\tau$
is an affine parameter:

\beqa 0 & = & {\partial \textsl{L}^2 \over \partial x^c} - {d
\over d \tau} {\partial \textsl{L}^2 \over \partial \dot{x}^c} =
\al_{,c} g_{ab} \dot{x}^a \dot{x}^b + \al g_{ab,c} \dot{x}^a
\dot{x}^b - {d \over d \tau} \left(\al^2 g_{ac} \dot{x}^a \right)
=\nonumber \\ & = & \left(\al_{,c} g_{ab} + \al g_{ab,c} \right)
\dot{x}^a \dot{x}^b -2 \dot{\al} g_{ac} \dot{x}^a - 2 \al g_{ac,b}
\dot{x}^a \dot{x}^b - 2 \al g_{ac} \dot{x}^a =\nonumber \\ & = &
\left({\al_{,c} \over \al} g_{ab} + g_{ab,c} \right) \dot{x}^a
\dot{x}^b - {2 \dot{\al} \over \al} g_{ac} \dot{x}^a - 2 g_{ac,b}
\dot{x}^a \dot{x}^b - 2 g_{ac}\ddot{x}^a =\nonumber \\ & = &
\left({\al_{,c} \over \al} g_{ab} + g_{ab,c} - 2g_{ac,b} \right)
\dot{x}^a \dot{x}^b - {2 \dot{\al} \over \al} g_{ac} \dot{x}^a - 2
g_{ac}\ddot{x}^a =\nonumber \\ & = & g_{ac} \dot{x}^a + \left[{1
\over 2} \left(g_{ac,b} + g_{bc,a} - g_{ab,c} \right) - {\al_{,c}
\over 2 \al} g_{ab} \right] \dot{x}^a \dot{x}^b + 2 {\dot{\al}
\over \al} g_{ac} \dot{x}^a =\nonumber \\ & = & g^{cd} g_{ab}
\ddot{x}^a + \left[{1 \over 2} g^{cd} \left(g_{ac,b} + g_{bc,a} -
g_{ab,c} \right) - {\al_{,c} \over 2 \al} g^{cd} g_{ab} \right]
\dot{x}^a \dot{x}^b + 2 {\dot{\al} \over \al} g^{cd} g_{ac}
\dot{x}^a =\nonumber \\ & = & \ddot{x}^d + \left[ \Gamma^d_{ab} -
{\al_d \over 2 \al} g_{ad} \right] \dot{x}^a \dot{x}^b + {2
\dot{\al} \over \al} \dot{x}^d \rightarrow \ddot{x}^a + \left[
\Gamma^a_{bc} - {\al_a \over 2 \al} g_{bc} \right] \dot{x}^b
\dot{x}^c + {2 \dot{\al} \over \al} \dot{x}^a = 0 ~~.\eeqa

In the isotropic, Newtonian case, one has $ \al = \al(r, \tau \sim
t)$, and thus

\beq \vec{a} = \vec{a}_N + {\al' \over 2 \al} g_{bc} \dot{x}^b
\dot{x}^c \vec{u}_r - {2 \dot{\al} \over \al} \vec{v} ~~, \eeq

\noindent where the prime denotes derivative with respect to the
radial coordinate.

Using $ g_{ab} = diag(1,-1,-1,-,1) $ so that $g_{bc} \dot{x}^b
\dot{x}^c = 1- v^2 \simeq 1$, one obtains a radial anomalous
acceleration plus a time-dependent drag force:

\beq a_A = {\al_{,r} \over 2\al} < 0, ~~ a_D = -{2 \dot{\al} \over
\al} < 0~~. \eeq

\vfill
%%%%%%%%%%%%%%%%%%%%%%%%%%%%%%%%%%%%%%%%%%%%%%%%%%%%%%%%%%%%%%%%
\newpage

\noindent {\bf Acknowledgments}

\vspace{0.2cm}

\noindent The authors wish to thank Urbano Fran\c{c}a, Il\'{i}dio
Lopes and Rog\'{e}rio Rosenfeld for useful discussions on the
solar astrophysics and on VAMP models. JP is sponsored by the
Funda\c{c}\~{a}o para a Ci\^{e}ncia e Tecnologia (Portuguese
Agency) under the grant BD~6207/2001.
%%%%%%%%%%%%%%%%%%%%%%%%%%%%%%%%%%%%%%%%%%%%%%%%%%%%%%%%%%%%%%%%

\end{document}